\documentclass[a4paper,11pt]{article}
\pdfoutput=1 

\usepackage{jcappub} 
                     
\usepackage[T1]{fontenc} 

\usepackage[english]{babel}
\usepackage{graphicx}
\usepackage{amsmath}
\usepackage{subcaption}
\usepackage{natbib}
\title{Forecast constraints on null tests of the $\Lambda$CDM model with SPHEREx}

\author[a]{A. Mata Rom\'an,}
\author[a]{I. Ocampo,}
\author[a,1]{S. Nesseris\note{Corresponding author.}}

\affiliation[a]{Instituto de F\'isica Te\'orica (IFT) UAM-CSIC, Calle Nicol\'as Cabrera 13-15, Campus de Cantoblanco UAM, 28049 Madrid, Spain}

\emailAdd{alejandro.matar@estudiante.uam.es}
\emailAdd{indira.ocampo@csic.es}
\emailAdd{savvas.nesseris@csic.es}

\abstract{In this work we quantify the ability  of the upcoming SPHEREx survey to constrain cosmological observables and test the internal consistency of the cosmological constant and cold dark matter ($\Lambda$CDM) model. Using Fisher matrix forecasting, we assess the expected precision on Baryon Acoustic Oscillations (BAO) observables, such as the angular diameter distance $D_\mathrm{A}(z)$ and the Hubble parameter $H(z)$. We further explore SPHEREx's potential to probe some of the fundamental assumptions of large-scale spatial homogeneity and isotropy, through model-independent reconstructions of several consistency tests of the $\Lambda$CDM model. In addition, we also examine the effect of the model dependence of the resulting Fisher and covariance matrices, using a neural network (NN) classification approach. We find that, while it is commonly assumed the covariance matrix depends weakly on the model, in fact the NN can very accurately ($\sim 98\%$) detect the underlying fiducial cosmological model based solely on the covariance matrix of the data, thus challenging this assumption. This model dependence, often neglected in standard analyses, can be naturally incorporated within simulation-based inference frameworks, which offer a flexible alternative for capturing such effects.}

\begin{document}
\maketitle
\flushbottom

\section{Introduction}
\label{sec:intro}
Recent advancements in observational techniques are driving cosmology into a new era of unprecedented data quality, enabling the precise testing of cosmological models. The current Stage IV cosmological surveys are already shaping the next decade of observational cosmology, delivering high-precision measurements and vast sky coverage. Among them, some flagship missions leading this endeavor are the Euclid satellite \cite{euclid_OV}, the Dark Energy Spectroscopic Instrument (DESI) \cite{DESI_OV}, the Large Synoptic Survey Telescope (LSST) \cite{LSST_book} and the SPHEREx mission \cite{SPX_OV}, but also the Vera C. Rubin Observatory \cite{VeraRubin} and the Nancy Grace Roman Space Telescope \cite{roman}. 

The SPHEREx mission, which is our focus in this work, is a NASA satellite designed to conduct the first all-sky spectral survey in the near infrared. Over its planned two-year lifetime, SPHEREx will map four times the entire celestial sphere, producing a three dimensional catalog of over 450 million galaxies along with more than 100 million stars. One of the key advantages of SPHEREx is its ability to reach redshift up to $z\sim6$, enabling studies that probe the age of reionization and the growth of the large-scale structure (LSS) across the cosmic history. A primary goal of SPHEREx is to constrain primordial non-Gaussianity by measuring galaxy clustering on ultra-large scales, allowing constraints on the local $f_\mathrm{NL}$ parameter with a sensitivity of $f_\mathrm{NL}\lesssim 1$. In addition, it will provide valuable data for studying the extragalactic background and tracing star formation over cosmic time. 

Overall, Stage IV surveys and SPHEREx in particular, offer a unique opportunity to test the standard cosmological constant $\Lambda$ and cold dark matter (CDM) model, known as $\Lambda$CDM. The latter can successfully explain current cosmological observations, but relies on two components whose nature is still unknown: dark matter (DM) and dark energy (DE). Additionally, $\Lambda$CDM still has some open problems: the Hubble tension \cite{Hubble_tension}, the $\sigma_8$ tension \cite{sigma8_tension}, the cosmic microwave background (CMB)  anisotropies anomaly, the cosmological constant problem \cite{cosmo_const_problem} and others \cite{Perivolaropoulos_2022}. 

As a result, modified gravity (MoG) models have been proposed, primarily to explain the late-time cosmic acceleration \cite{Woodard_2007} and structure formation \cite{Tsujikawa_2007}. This includes, for instance, $f(R)$ theories \cite{fR_theory} and higher-order gravity models \cite{Bogdanos_2010} among others. However, any modification to gravity must satisfy solar system and laboratory tests. This is achieved through screening mechanisms, which suppress deviations from general relativity (GR) in high-density environments while allowing modifications to appear on large scales \cite{Hinterbichler_2010}. Examples of screening mechanisms include the chameleon, Vainshtein and K-mouflage mechanisms, among others \cite{chamaleon}. 

Moreover, Stage IV surveys are not only essential to constrain cosmological parameters but also for testing the consistency of these models by different observational probes, such as BAO or weak lensing. There have been already several forecasts on the performance of these surveys \cite{casas2023euclidconstraintsfrcosmologies,Nesseris_2022,Euclid_forecast,Bernal_2019,shiveshwarkar2023postinflationarycontaminationlocalprimordial,dinda2025improvednulltestslambdacdm,Dinda_2024}. Recent results from DESI indicate that the DE equation of state parameters $w_0$ and $w_a$, representing the present day value of the DE equation of state and its first derivative respectively, differ from the expected values of the cosmological constant model of $(w_0,w_a)=(-1,0)$ \cite{DESI:2025zgx}, underscoring the importance of Stage IV surveys.

In this context, we then focus on the SPHEREx mission to test the consistency of the spatially flat $\Lambda$CDM model, taking advantage of its unprecedented redshift coverage. We use Fisher matrix forecasting to predict how precisely SPHEREx will measure key cosmological parameters and machine learning algorithms to detect any deviation on the spatial homogeneity and isotropy of the Universe via null tests, i.e. consistency tests, of the $\Lambda$CDM model. 

In addition, we present a neural network approach aiming to test the common assumption that the covariance and Fisher matrices depend weakly on the assumed cosmological model, by performing classification between the $\Lambda$CDM and $w$CDM models. Thus, with our method we aim to quantify the extent of model dependence and assess whether machine learning can effectively detect these subtle differences from different cosmological scenarios. This model and parameter dependence of the Fisher matrix is an important aspect often overlooked in standard analyses that rely on a single fiducial covariance. Our findings highlight the need for complementary approaches that can naturally account for such dependencies, such as simulation-based inference (SBI) or likelihood-free methods \cite{Cranmer2020,Alsing2019,Charnock2018}, which learn the full mapping between cosmological parameters and observables directly from simulations without assuming a fixed local approximation.

\section{Theoretical framework}
\subsection{Cosmological background}
The Cosmological Principle (CP) states that the Universe is spatially homogeneous and isotropic at sufficiently large scales, i.e. distances larger than $\sim 100$\,Mpc \cite{Scrimgeour:2012wt}. Under this assumption, the spacetime geometry must exhibit maximal symmetry in its spatial sections, leading uniquely to the Friedman-Lema\^itre-Robertson-Walker (FLRW) metric,
\begin{equation}
    ds^2=-dt^2+a^2(t)
    \left[\,\frac{dr^2}{1-K\,r^2}+r^2\,(d\theta^2+\sin^2\theta \,d\phi^2)\,\right],
\end{equation}
where the scale factor $a(t)$ accounts for the expansion of the Universe, while $K$ is the curvature parameter.  

Inserting this metric into Einstein's field equations and assuming a perfect fluid with energy density $\rho$ and pressure $P$, the dynamics of the expansion of the Universe are governed by the Friedman equations. In the presence of multiple components, the total energy density is given by $\rho=\sum_i\rho_i$, where each component, assuming a barotropic fluid, is characterized by its equation-of-state (EoS) parameter $w_i=P_i/\rho_i$. 

Assuming spatial flatness, i.e. $K=0$, and neglecting radiation and neutrinos at late times, we can express the Friedmann equation as follows
\begin{equation}
    H^2(a)/ H_0^2\equiv E^2(a)=\Omega_{\mathrm{m},0}\,a^{-3}+\Omega_\mathrm{DE,0}\;a^{-3\;(1+w)}\,,
    \label{H_DE}
\end{equation}
where we impose the consistency relation $\Omega_{\mathrm{m},0}+\Omega_\mathrm{DE,0}=1$, $H(a)$ is the Hubble parameter, $H_0=100\,h\, \mathrm{km}\,\mathrm{s^{-1}\,\mathrm{Mpc}^{-1}}$ is the present day value of the Hubble parameter, $\Omega_i=\rho_{i,0}/\rho_\mathrm{cr,0}$ is the present-day density parameter of component $i$ and $\rho_\mathrm{cr,0}=3\,H_0^2/(8\pi G_\mathrm{N})$ and $G_\mathrm{N}$ is Newton's constant. Note that at late times we can ignore any contribution coming from radiation and neutrinos, thus we have set $\Omega_{\mathrm{r},0}=0$ and $\Omega_{\nu,0}=0$. Furthermore, in some cases, it will be more convenient to use the redshift $z$ instead of the scale factor $a$, related to each other via $a=(1+z)^{-1}$ \cite{Dodelson}. 

The $\Lambda$CDM model provides the simplest and most successful description of the Universe, in which the DE component corresponds to a cosmological constant ($\Lambda$) with $w=-1$. However, a widely studied phenomenological extension from $\Lambda$CDM is the $w$CDM model, where the DE EoS is still assumed to be constant, but not necessarily equal to -1. In this context, different values of $w$ lead to different cosmological scenarios. For example, if $w>-1/3$, the DE pressure would not be negative enough to overcome the gravitational attraction and the expansion of the Universe would be decelerated, which is not compatible with observations. Therefore, viable cosmological models require $w<-1/3$. 

The regime  $-1\leq w<-1/3$ corresponds to the quintessence scenario, where DE arises from a slowly evolving scalar field minimally coupled to gravity \cite{Tsujikawa_quintess}. Quintessence models can be further classified, for instance, tracking or scaling freezing models. Conversely, for $w<-1$, DE falls into the phantom regime \cite{Nesseris_phantom}, characterized by an excessive negative pressure that leads to a super-accelerated expansion, such as the Big Rip \cite{caldwell2003phantom, Nesseris:2004uj,TrivediScherrer2024}. Despite leading to some catastrophic scenarios, $w<-1$ is not yet ruled out.

An important feature for our forecast are distances. In cosmology there are multiple definitions of distance depending on how we measure it, e.g. the comoving distance measures the `coordinate distance' traveled by a photon from redshift $z$, given by
\begin{equation}
    r_\mathrm{c}(z)=\int \frac{dt}{a(t)}=\frac{c}{H_0}\int_0^z \frac{dx}{E(x)}\;.
    \label{cov_distance}
\end{equation}

The comoving distance is not a direct observable, but it is linked to other distance definitions related to observables. For instance, the angular diameter distance, $D_\mathrm{A}(z)=r_\mathrm{c}(z)\,(1+z)^{-1}$, which represents the ratio between the source's physical transverse size and its observed angular size. Conversely, the luminosity distance, $D_\mathrm{L}(z)=(1+z)\,r_\mathrm{c}(z)$, which connects the intrinsic luminosity and the observed flux. The last two quantities define the distance duality relation
\begin{equation}
    \frac{D_\mathrm{L}(z)}{D_\mathrm{A}(z)\,(1+z)^2}=1\;,
\end{equation}
which holds under two main conditions: photons travel along the null geodesics, and their number is conserved, thus any deviation from unity would imply new physics or unaccounted astrophysical effects.

\subsection{Consistency tests} \label{sec::consist_sect}
Among different approaches to test cosmological theories, model-agnostic consistency tests, also known as null tests, stand out for their simplicity and effectiveness. These tests are constructed in such a way that they do not depend on fitting a specific model's parameters, but rather on defining quantities that are constant at all redshifts. In other words, these tests involve defining functions of observables that should yield a fixed value, if the underlying hypotheses are true.  

The reason for this is that in a spatially homogeneous and isotropic Universe, background quantities can only depend on time, or equivalently, the redshift $z$. Since we are interested in using background quantities, such as the Hubble expansion rate or cosmological distances, then we can use the latter to create null tests, i.e. pass-fail tests, that have to be constant in redshift, as this is the only possible ``time'' variable. Then, one of the easiest ways to create these tests is to define constant quantities via combinations of background variables and, given some model agnostic reconstructions of the latter, to probe for any deviations from a constant value at any redshift.

Thus, it is essential to carefully select consistency tests that can effectively examine these assumptions. In what follows, we focus on forecasting how well some of these tests can constrain deviations from the $\Lambda$CDM model, using future surveys such as SPHEREx.  

\subsubsection{The $O\mathrm{m}_\mathrm{H}$ diagnostic}
Using Eq.~\eqref{H_DE}, with $w=-1$ and assuming spatial flatness, but also neglecting neutrinos and radiation at late times, we can solve the Friedmann equation for $\Omega_{\mathrm{m},0}$ and define the quantity $O\mathrm{m}_\mathrm{H}$ which goes to the constant $\Omega_{\mathrm{m},0}$ value if and only if $\Lambda$CDM is the true model \cite{Sahni_2008}:
\begin{equation}\label{eq:om_statistic}
    O\mathrm{m}_\mathrm{H}(z)=\frac{E(z)^2-1}{(1+z)^3-1}\;\rightarrow \Omega_{\mathrm{m},0}~\mathrm{(only~in~}\Lambda\mathrm{CDM)}.
\end{equation}

The $O\mathrm{m}_\mathrm{H}$ test can be seen as a model-independent assessment of the $\Lambda$CDM expansion history, as it directly probes the functional form of $H(z)$ without requiring DE assumptions. 
However, an issue with this diagnostic is that it is ill-defined when $z\to0$, where the denominator vanishes. If the uncertainties in both numerator and denominator do not scale in a comparable way near this limit, the resulting $O\mathrm{m}_\mathrm{H}$ may be systematically biased, reducing the test reliability at small redshifts. While we could regularize the singularity at small redshifts by multiplying $O\mathrm{m}_\mathrm{H}$ by $z$, for historical reasons and in order to compare our results with those of the literature, we do not do so.

Prior reconstructions of the $O\mathrm{m}_\mathrm{H}$ diagnostic tests have been already explored \cite{Nesseris_2010,Nesseris_2022}. For instance, the Euclid collaboration used this diagnostic with external and forecasted data to test deviations from $\Lambda$CDM, showing that \textit{Euclid} will be able to constrain such deviations at a few percent level across a broad redshift range \cite{Nesseris_2022}. 

\subsubsection{The curvature test}
The comoving distance defined in Eq.~\eqref{cov_distance} can be expressed in a more general form following
\begin{equation}
    r_\mathrm{c}(z)=\frac{c}{H_0}\,\frac{1}{\;\sqrt{-\Omega_{k,0}}}\;
    \sin\left(\sqrt{-\Omega_{k,0}}\;\int_0^z \frac{dx}{E(x)}\right)\;,
    \label{cov_distance_k}
\end{equation}
where $\Omega_{k,0}$ is the curvature parameter today. If we solve for $\Omega_{k,0}$, we find an expression that depends uniquely on $E(z)$ and $\tilde{r}_\mathrm{c}(z)=\frac{H_0}{c}\,r_\mathrm{c}(z)$,
known as the curvature test \cite{Clarkson_Oktest}
\begin{equation}\label{eq:ok_test}
    \Omega_k(z)=\frac{[\;E(z)\;\tilde{r}_\mathrm{c}'(z)\;]^2-1}{\tilde{r}_\mathrm{c}^2(z)}\;,
\end{equation}
where $\tilde{r}_\mathrm{c}'(z)=\partial_z\,\tilde{r}_\mathrm{c}(z)$. The aforementioned expression not only can be used to test the curvature of the Universe, but it provides a unique way of measuring the current curvature parameters by means of the Hubble rate and the comoving distance.  Importantly, it remains constant for any redshift regardless of the DE model assumed, making it a powerful probe not just of curvature, but also of the large-scale spatial homogeneity of the Universe, as in a spatially homogeneous Universe, the curvature has to be constant everywhere.  

Nevertheless, the curvature test diverges as $\sim1/z^2$ at low redshift \cite{Sapone_2014}. By considering the quantity $z^2\,\Omega_k$, the factor $z^2$ cancels out the divergent behavior, resulting in a well-behaved quantity at low redshift. This rescaling provides a more stable and meaningful diagnostic near $z\sim0$. However, the uncertainties at higher redshift will scale accordingly, requiring careful interpretation.

\subsubsection{Global shear test}
A direct test of the large-scale spatial isotropy of the Universe is the normalized global shear
\begin{equation}
    \Sigma(z)\equiv\frac{H_\mathrm{T}-H_\mathrm{L}}{H_\mathrm{L}+2\;H_\mathrm{T}},
    \label{global_shear1}
\end{equation}
where $H_\mathrm{T}$ and $H_\mathrm{L}$ represent the transversal and longitudinal Hubble expansion rates in the Lema\^itre-Tolman-Bondi (CLP) metric respectively \cite{Garc_a_Bellido_2009}. This dimensionless quantity measures any potential difference in the expansion rates along the radial and angular directions. In an isotropic universe, the Hubble parameter  must evolve accordingly in all directions, i.e. $H_\mathrm{T}=H_\mathrm{L}$. Consequently, the global shear would vanish in any FLRW model, independent of the curvature. Thereafter, any deviation from $\Sigma(z)=0$ would indicate anisotropies in the cosmic expansion, challenging the Copernican Principle.  

Assuming zero curvature and rewriting Eq.~\eqref{global_shear1} in terms of the dimensionless Hubble parameter $E(z)$ and the dimensionless comoving distance $\tilde{r}_\mathrm{c}(z)$, the global shear takes the form
\begin{equation}
    \Sigma(z)= \frac{1-E(z)\;\tilde{r}_\mathrm{c}'(z)}{3\;\frac{H_0}{c}\;E(z)\;D_\mathrm{A}(z)+2\;[1-E(z)\;\tilde{r}_\mathrm{c}'(z)]\;}\;.
    \label{global_shear2}
\end{equation}

\subsubsection{Other tests}
While there are several more null tests that probe for deviations from the flat $\Lambda$CDM model, for example the extended $\mathrm{Om}_H(z)$ statistic with curvature \cite{Seikel:2012cs}, the $r_0$ test for interactions in the dark sector \cite{vonMarttens:2018bvz,vonMarttens:2020apn}, the distance null tests \cite{Arjona:2019fwb} and the various growth-rate data null tests: the growth-index $\gamma$ \cite{Wang:1998gt}, the $\mathcal{O}(a)$ null test \cite{Nesseris:2014mfa}, and the $\mathrm{Om}_{f\sigma_8}(a)$ test \cite{Arjona:2021mzf}, in this work we focus only on the tests mentioned in the previous subsections in order to keep the analysis clear and simple, and we leave a comparison of all the tests for future work. 

\subsection{Fisher matrix formalism}
The Fisher matrix formalism is a statistical method that allows us to easily forecast constraining power on cosmological parameters. In other words, the Fisher matrix measures the amount of information that an observable carries about a specific parameter. It is defined as the expected value of the second derivative of the logarithmic likelihood evaluated at a reference parameter value $\theta_\mathrm{ref}$, corresponding to a fiducial model under consideration. Specifically, it is given by
\begin{equation}
    F_{\alpha\beta}(\theta)=\left<\left.-\frac{\partial^2\ln L}{\partial\theta_\alpha\,\partial\theta_\beta}\right|_{\theta_\mathrm{ref}}\right>\;,
\end{equation}
where $\alpha$ and $\beta$ label the parameters $\theta_\alpha$ and $\theta_\beta$. 

Once we have computed the Fisher matrix, the covariance matrix of the parameters is just the inverse of the Fisher matrix
$C_{\alpha\beta}=\left[F_{\alpha\beta}(\theta)\right]^{-1}$. The marginalized $68.3\%$ credible region (corresponding to the 1$\sigma$ in the Gaussian case) are encoded in the diagonal terms of the matrix such that $\sigma_{\alpha}=\sqrt{C_{\alpha\alpha}}$ and the non-diagonal terms represent the correlation between two parameters, i.e. $C_{\alpha\beta}=\rho_{\alpha\beta}\,\sigma_\alpha\,\sigma_\beta$. The correlation parameter $\rho_{\alpha\beta}$ quantifies the degree of linear dependence between the parameters and is bounded in the range $[-1,1]$, as a consequence of the Cauchy-Schwarz inequality \cite{cauchySchwarz}. For instance, if two parameters are perfectly correlated, then $\rho_{\alpha\beta}=1$, whereas perfect anticorrelation corresponds to $\rho_{\alpha\beta}=-1$.

The marginalized Fisher matrix for a subset of parameters is the inverse of the covariance matrix restricted to those parameters. Thereafter, to marginalize over some parameters we just need to remove from the covariance matrix the rows and columns related to those parameters and invert the resulting covariance matrix to obtain the new Fisher matrix \cite{Amendola_Tsujikawa_2010}. 

Suppose that we have constructed our Fisher matrix using a set of parameters $\boldsymbol{\theta}$ but we want to construct a new Fisher matrix based on a different set of parameters $\boldsymbol{\phi}$. To compute the 1$\sigma$ credible region of this new parameter space, a common procedure is to do error propagation with the covariance matrix after marginalizing over nuisance parameters. If we denote the reduced density matrix as $\widetilde{C}_{\alpha\beta}$, then the covariance matrix $S_{ij}=F_{ij}^{-1}(\phi)$ of the parameters in the $\boldsymbol{\phi}$ basis is \footnote{Note that in our work Greek letters like $\alpha$, $\beta$ correspond to parameters in the $\boldsymbol{\theta}$ basis, while Latin letters like $i, j$ correspond to the $\boldsymbol{\phi}$ basis.  }
\begin{equation}
    S_{ij}(\boldsymbol{\phi})=\sum_{\alpha\beta}\,
    \widetilde{C}_{\alpha\beta}\,\frac{d\phi_i}{d\theta_\alpha}\,\frac{d\phi_j}{d\theta_\beta}\;.
    \label{err_prop}
\end{equation}

However, these transformations can lead to a non-positive-definite Fisher matrix, for instance, due to numerical errors and the high correlation between the bins. This can happen due to the fact that when the correlation coefficient $\rho$ is sufficiently close to $\pm 1$, then the inverse of the covariance matrix may be singular, as its determinant will be (numerically) very close to zero (obviously having a non-zero determinant is a condition for a matrix to be invertible). For example, for two random variables with errors $\sigma_1$ and $\sigma_2$, this would correspond to having a covariance matrix 
$$
C_{ij=}\begin{pmatrix}
\sigma_1^2 & \rho\,\sigma_1\,\sigma_2 \\
\rho\,\sigma_1\,\sigma_2 & \sigma_2^2 
\end{pmatrix},
$$
that has a determinant given by $|C|=(1-\rho^2)\,\sigma_1^2\,\sigma_2^2$, which for $\rho\simeq \pm1$ is also approximately equal (within machine precision) to zero. In the latter case then its inverse $F_{ij}=C_{ij}^{-1}$ can be numerically unstable and singular. 

Furthermore, this is problematic because the chi-squared, defined as $\chi^2=X^i\,F_{ij}(\boldsymbol{\phi})\,X^j$, will not then be positive for all non-zero $\mathbf{X}$. To address this issue, a common procedure consists of shifting the eigenvalues of the matrix to eliminate any negative components, ensuring positive definiteness. This can be done by redefining the Fisher matrix as \cite{eigenshift}
\begin{equation}
    F'_{ij}(\boldsymbol{\phi})=F_{ij}(\boldsymbol{\phi})+\delta_{ij}\,(1+|\alpha|)\,|\lambda_{\text{min}}|\;,
    \label{fish_curing}
\end{equation}
where $\lambda_{\text{min}}$ is the smallest, i.e. most negative, eigenvalue and $\alpha$, some parameter that controls the magnitude of the shift. It should be stressed that adding a constant to the Fisher matrix before inverting has the interpretation of applying a tighter prior on the posterior forecast, however we have taken steps (discussed later in Section 4.1), to make sure our results are robust. 

\section{Fisher matrix for the power spectrum}
In order to forecast the sensitivity of an LSS survey, such as SPHEREx, we will now construct our Fisher matrix in terms of the observed power spectrum $P(k,\mu;z)$, where $k$ is the modulus of the wave vector in Fourier space, and $\mu=\cos\theta$ is the cosine of the angle between the wave vector and the line-of-sight, i.e. $\theta=\mathbf{\hat{k}\cdot\mathbf{\hat{r}}}$. The power spectrum, defined as the Fourier transform of the two-point correlation function $P_{\delta\delta}=\mathcal{F}[\xi(\mathbf{r},z)]$, is a natural choice for this analysis, as it quantifies the excess probability of finding two galaxies separated by a given distance relative to a random distribution, thereby capturing the degree of matter clustering in the Universe across different scales. Nevertheless, we still need to apply some observational effects that may distort its full shape.  

\subsection{The observed power spectrum}
Observationally, we can only account for the baryonic matter present in galaxies, but we are ultimately interested in the CDM distribution. To bridge this gap, we introduce the galaxy bias $b$, which relates the CDM and matter baryonic matter distributions. The biased power spectrum is then expressed as $P_b(k;z)=b^2\,P_{\delta\delta}(k;z)$. If $b<1$ it means that galaxies are less clustered than the underlying CDM distribution, whereas $b>1$ implies they are more clustered. The bias parameter generally increases with redshift, as galaxies at earlier times tend to reside in higher density peaks of matter distribution. Additionally, on small scales, $b$ exhibits scale dependence due to the influence of complex processes such as halo exclusion and baryonic effects. In contrast, on large scales, the bias approaches a scale-independent form. Further discussion can be found at \cite{Desjacques_2018}. 

Due to peculiar galaxy velocities, distortions arise when converting redshifts into sky positions \cite{Dodelson}. For instance, a slightly overdense region would appear elongated along the line-of-sight,  which in the linear regime is called the Kaiser effect \cite{kaiser} and can be modeled following
\begin{equation}
    P_\mathrm{kaiser}(k,\mu;z)=\left[b(z)\,\sigma_8(z)+f(z)\,\sigma_8(z)\mu^2\right]^2
    \,\frac{P_{\delta\delta}(k;z)}{\sigma_8^2(z)}\;,
    \label{kaiser_eq}
\end{equation}
where $\sigma_8(z)=\sigma_{8,0}\,\delta(z)/\delta(0)$ is the root means square (RMS) density fluctuation at a redshift $z$, $\sigma_{8,0}$ is its present day value, $\delta(a)=\delta \rho_\mathrm{m}/\rho_\mathrm{m}$ is the matter density contrast and $f(a)=\mathrm{d}\ln \delta/\mathrm{d}\ln a$ is its logarithmic derivative, called the growth-rate.
The $\sigma_{8,0}$ parameter is explicitly given by \cite{Dodelson}
\begin{equation}
\sigma_8^2 = \frac{1}{2\pi^2} \int_0^{\infty} k^2 P_{\delta\delta}(k,z=0)\, W^2(kR)\, \mathrm{d}k,
\end{equation}
where $W(kR)$ is a window function, typically assumed to be a top-hat 
\begin{equation}
    W(kR) = \frac{3[\sin(kR) - kR\cos(kR)]}{(kR)^3}.
\end{equation}

In the radial direction, the uncertainty in the redshift suppresses the observed power spectrum at large redshift. This can be taken into account by adding an exponential term $F(k,\mu;z)=\text{exp}[-k^2\,\mu^2\,\sigma^2_r(z)]$ such that
\begin{equation}
    P_{z,\mathrm{lin}}(k,\mu;z)= P_\mathrm{kaiser}(k,\mu;z)\,F_z(k,\mu;z)\;,
\end{equation}
where $\sigma_r(z)=c\,(1+z)\,\sigma_{0,z}\,/\,H(z)$ and $\sigma_{0,z}$ is the redshift uncertainty. 

Reconstructing the power spectrum also requires a reference cosmology to convert redshifts to distances. These discrepancies between the fiducial and true cosmology induce a systematic known as the Alcock-Paczynski (AP) effect \cite{alcock}. To obtain the power spectrum in the true cosmology we need to preserve the redshift and angular diameters of the observed galaxies. This can be done by taking the ratios
\begin{equation}
    q_\perp(z)=\frac{D_\mathrm{A}(z)}{D_\mathrm{A,ref}(z)} \;,\;\;\;\;\;\;\;\;\;\;q_\parallel(z)=\frac{H_\mathrm{ref}(z)}{H(z)}\;,
\end{equation}
and redefining the true $k$ and $\mu$ as
\begin{equation}
    k_\mathrm{AP}= \frac{k_\mathrm{ref}}{q_\perp}\,\left[ 1+\mu^2_\mathrm{ref}\,\left(\frac{q_\perp}{q_\parallel}-1\right)\right]^{1/2}\;,
\end{equation}
\begin{equation}
    \mu_\mathrm{AP}= \mu^2_\mathrm{ref}\,\frac{q_\perp}{q_\parallel}\,\left[ 1+\mu^2_\mathrm{ref}\,\left(\frac{q_\perp}{q_\parallel}-1\right)\right]^{-1/2}\;,
\end{equation}
where $k_\mathrm{ref}$ and $\mu_\mathrm{ref}$ are the ones used in our reference cosmology. For notational convenience, we hereafter denote $k=k_\mathrm{AP}$ and $\mu=\mu_\mathrm{AP}$. Incorporating the AP effect, the resulting power spectrum is
\begin{equation}
    P_\mathrm{obs,lin}(k,\mu;z)=\frac{1}{q_\perp^2 \,q_\parallel}\,P_{z,\mathrm{lin}}(k,\mu;z)\;.
\end{equation}
Accurately modeling the observed power spectrum requires accounting for non-linear effects that significantly alter its shape at small and large scales. To characterize the impact of these effects, we introduce two phenomenological parameters $\sigma_p(z)$ and $\sigma_v(z)$ which describe the linear galaxy velocity dispersion \cite{Euclid_forecast}
\begin{equation}
    \sigma_p^2(z)=\sigma_v^2(z)=\frac{1}{6\,\pi^2}\int dk\,P_{\delta\delta}(k;z)\;.
\end{equation}
There are two particular non-linear phenomena relevant here. First, the Kaiser effect introduced in \eqref{kaiser_eq} does not incorporate the so-called Finger of God (FoG) effect. This effect is usually modeled by a Lorentzian damping factor. Second, the non-linear evolution of clustering progressively smooths out the characteristic BAO features, damping its signal in both transversal and longitudinal directions. 

To address this, it is necessary to incorporate a model that captures the gradual suppression of BAO oscillations while preserving the broadband shape of the power spectrum. The so-called `de-wiggled' power spectrum $P_\mathrm{dw}(k,\mu;z)$ \cite{Euclid_forecast} provides a solution to this issue. This model interpolates smoothly between the original linear power spectrum $P_{\delta\delta}$ and a wiggle-free version $P_\mathrm{nw}$, where the non-linear damping has erased the acoustic features. The transition is controlled by an exponential damping factor that captures the broadening of the BAO peaks induced by non-linear velocity dispersions and the growth of structure. Explicitly, the `de-wiggled' power spectrum is given by
\begin{equation}
    P_\mathrm{dw}(k,\mu;z)=P_{\delta\delta}(k,\mu;z)\,e^{-g_\mu k^2}+
    P_\mathrm{nw}(k,\mu;z)\,(1-e^{-g_\mu k^2})\;,
\end{equation}
where $g_{\mu}(k,\mu,z)=\sigma_v(z)\left\{1-\mu^2+\mu^2\,[\,1+f(z)\,]^2\right\}$ quantifies the anisotropic damping of the BAO. Then, taking into account all observational effects, the full non-linear observed power spectrum is \cite{Euclid_forecast}
\begin{equation}
    P_\mathrm{obs}(k,\mu;z)= \frac{1}{q_\perp^2\, q_\parallel}\frac{\left[\,b(z)\,\sigma_8(z)+f(z)\,\sigma_8(z)\,\mu^2\,\right]^2}{1+[\,f(z\,)k\,\mu\,\sigma_p(z)\,]^2} \frac{P_\mathrm{dw}(k,\mu;z)}{\sigma_8^2(z)}\,F_z(k,\mu;z)\;.
    \label{pk_obs}
\end{equation}

Once we have our full power spectrum, we can compute the Fisher matrix in each redshift bin as \cite{Amendola_Tsujikawa_2010}
\begin{equation}
    F_{\alpha\beta}^\mathrm{bin}(\theta,z_i)=
    \frac{1}{8\pi^2}\int_{-1}^1 d\mu \int_{k_\mathrm{min}}^{k_\mathrm{max}} dk\;k^2\;
    \frac{\partial\ln P_\mathrm{obs}(k,\mu;z_i)}{\partial\theta_\alpha}\;
    \frac{\partial\ln P_\mathrm{obs}(k,\mu;z_i)}{\partial\theta_\beta}\;V_\mathrm{eff}\;,
    \label{Fish_eq}
\end{equation}
where the integral over $k$ arises from the summation over all independent Fourier modes available within the survey volume. The limits $k_\mathrm{min}$ and $k_\mathrm{max}$ correspond to the largest and smallest scales covered by the survey. The comoving survey volume between two redshifts $z_i$ and $z_f$ over a solid angle $\Omega$ is given by,
\begin{equation} 
V_s(z_i,z_f)=\Omega\int_{r_\mathrm{c}(z_i)}^{r_\mathrm{c}(z_f)} r_\mathrm{c}^2 \,dr=\frac{\Omega}{3}\,\left[r_\mathrm{c}^3(z_f)-r_\mathrm{c}^3(z_i)\right]\;, \label{cov_vol} 
\end{equation}
where $r_c(z)$ is the comoving distance. As is standard practice, we evaluate all quantities at the effective redshift $z_\mathrm{eff}$ of each bin, following \cite{Euclid_forecast}. The corresponding effective volume, which determines the statistical weight of modes contributing to the measurement, is then,
\begin{equation}
    V_\mathrm{eff}=V_s(z)\,\left[\,\frac{n(z)\,P(k,\mu;z)}{1+n(z)\,P(k,\mu;z)}\,\right]^2\;,
\end{equation}
where $V_s(z)$ is the survey volume corresponding to the effective redshift bin $z_\mathrm{eff}$, following \eqref{cov_vol}, and  $n(z)$ denotes the number density of galaxies within that bin. The effective volume quantifies the fraction of the survey volume that contributes to the measurement of the power spectrum at a given scale and orientation. In particular, the signal-to-noise ratio per Fourier mode is determined by the product $n(z)\,P(k,\mu;z)$. When $n(z)\,P(k,\mu;z)\gg1$, the measurement is signal dominated, and the effective volume approaches the total survey volume in that redshift bin, i.e. $V_\mathrm{eff}(z)\sim V_s(z)$. 

On the other hand, when $n(z)\,P(k,\mu;z)\ll1$, the signal becomes noise-dominated due to the low galaxy density, and the effective volume becomes negligible. Then, the full Fisher matrix can be computed summing over all redshift bins,
\begin{equation}
    F_{\alpha\beta}(\theta)=\sum_i^{N}\,F_{\alpha\beta}^\mathrm{bin}(\theta,z_i)\;.
    \label{fish_sum}
\end{equation}

\subsection{Survey specifications}
In this section, we present the SPHEREx survey specifications adopted for our Fisher forecasting. Specifically, SPHEREx is designed as an all-sky survey with a sky coverage of $f_{\text{sky}}=0.65$, covering scales between $k_\mathrm{min}=0.001\;h \,\mathrm{Mpc}^{-1}$ and $k_\mathrm{max}=0.2\;h\,\mathrm{Mpc}^{-1}$ \cite{SPX_OV}. This assumption is well justified on the large scales probed in our analysis. For wavenumbers up to $k_{\max}=0.2\,h\,\mathrm{Mpc}^{-1}$, non-linear and higher-order bias contributions are expected to have a subdominant impact on the overall forecast precision, especially given that Fisher forecasts primarily capture relative parameter sensitivities rather than detailed amplitude effects. Nonetheless, for analyses extending to smaller scales, a more accurate treatment including higher order corrections to the bias term would be required, which we leave for future work. 

A key advantage of SPHEREx over other contemporary redshift surveys is that galaxies are classified into different samples with different redshift uncertainties, galaxy biases, and number densities. This enables a more refined modeling of the observational effects depending on the main objective of the analysis \cite{SPX_OV}.

Specifically, SPHEREx provides five distinct galaxy samples, each divided into eleven redshift bins and with their associated redshift uncertainty \cite{SPX_sigma}. Although the sample with the lowest redshift uncertainty, $\sigma/(1+z)<0.003$, is best suited for resolving the BAO peaks in the radial direction, the other samples still contain angular and radial information, but with a higher redshift uncertainty.\footnote{Private communication with Prof. Olivier Doré, the Head Project Scientist of SPHEREx.} 

Thereafter, considering additional galaxy samples, with higher redshift uncertainty, would increase the number of galaxies, and thus the statistical precision and information. However, in this work, we focus specifically on galaxy clustering and the measurement of the BAO using the H$_\alpha$ emission line, which requires precise redshift measurements to resolve the clustering signal. For this reason, we restrict our analysis to the sample with the best available redshift precision, i.e. $\sigma/(1+z)\leq0.003$. Table~\ref{spx_spec} summarizes the data from the selected sample, while the full specifications of this sample are available online.\footnote{\label{spx_foot}\url{https://github.com/SPHEREx/Public-products/blob/master/galaxy_density_v28_base_cbe.txt}}  

\begin{table}[!t]
    \centering
    \begin{tabular}{c c c c c} \hline
        $z_{\text{min}}$ & $z_{\text{max}}$ & $n(z_{\text{mean}})\;[\;h^3\;\text{Mpc}^{-3}\;]$ & $V_s(z_\text{mean})\;[\;\text{Gpc}^{3}\;h^{-3}\;]$ & $b(z_\text{mean})$ \\ \hline\hline
         0.0 & 0.2 & $9.97\cdot10^{-3}$ & 1.68 & 1.3 \\ \hline
         0.2 & 0.4 & $4.11\cdot10^{-3}$ & 9.72 & 1.5 \\ \hline
         0.4 & 0.6 & $5.01\cdot10^{-4}$ & 21.15 & 1.8 \\ \hline
         0.6 & 0.8 & $7.05\cdot10^{-5}$ & 32.71 & 2.3 \\ \hline
         0.8 & 1.0 & $3.16\cdot10^{-5}$ & 42.82 & 2.1 \\ \hline
         1.0 & 1.6 & $1.64\cdot10^{-5}$ & 170.05 & 2.7 \\ \hline
         1.6 & 2.2 & $3.59\cdot10^{-6}$ & 200.25 & 3.6 \\ \hline
         2.2 & 2.8 & $8.07\cdot10^{-7}$ & 205.96 & 2.3 \\ \hline
         2.8 & 3.4 & $1.84\cdot10^{-6}$ & 200.54 & 3.2 \\ \hline
         3.4 & 4.0 & $1.50\cdot10^{-6}$ & 190.59 & 2.7 \\ \hline
         4.0 & 4.6 & $1.13\cdot10^{-6}$ & 179.15 & 3.8 \\ \hline
    \end{tabular}
    \caption{The SPHEREx survey specifications.\footref{spx_foot}}
    \label{spx_spec}
\end{table}

\section{Methodology} \label{meth_sect}
\subsection{Fisher forecast}
For our Fisher forecast analysis, we adopt a flat $\Lambda$CDM cosmological model with the following set of parameters $\boldsymbol{\theta}=\{\Omega_{\mathrm{m},0},\;\Omega_{\mathrm{b},0},\;h,\;n_s,\;\sigma_8\}$,
whose fiducial values are summarized in Table~\ref{fid_params}.

\begin{table}[!t]
    \centering
    \begin{tabular}{c c c c c} \hline
        $\Omega_{\mathrm{m},0}$ & $\Omega_{\mathrm{b},0}$ & $h$ & $n_s$ & $\sigma_8$ \\ \hline\hline
         0.32 & 0.05 & 0.67 & 0.96 & 0.816 \\ \hline
    \end{tabular}
    \caption{Parameter values for the $\Lambda$CDM fiducial model used.
    }
    \label{fid_params}
\end{table}

The Fisher matrix has been computed numerically using Eqs.~\eqref{Fish_eq} and \eqref{fish_sum}, while the derivatives of the power spectrum $P_\mathrm{obs}(k)$, defined in Eq.~\eqref{pk_obs}, were estimated using the central finite difference method, according to 
\begin{equation}
    \frac{\partial P_\mathrm{obs}(k,\mu;z_j)}{\partial\theta_i}=
    \frac{ P_\mathrm{obs}(k,\mu;z_j\;|\;\theta_i+\epsilon)-P_\mathrm{obs}(k,\mu;z_j\;|\;\theta_i-\epsilon)}{2\;\epsilon}\;,
\end{equation}
where $\epsilon=\eta\;\theta_i$, with $\eta\ll1$, is a small perturbation. We have explicitly verified that the estimation of derivatives using this finite-difference scheme is numerically stable with respect to variations of $\eta$, ensuring the robustness of the method. Finally, the power spectrum was computed using the Boltzmann solver \texttt{CLASS} \cite{CLASS_Blas}.

However, for the purpose of our consistency tests, we are primarily interested in a different parameter space defined by $\boldsymbol{\phi}=\{H(z_1),D_\mathrm{A}(z_1),\ldots, H(z_{11}),D_\mathrm{A}(z_{11})\}$, where $z_j$ corresponds to each of the SPHEREx redshift bins. To change the cosmological basis, we perform a projection from $\boldsymbol{\theta}\to\boldsymbol{\phi}$ using the error propagation formalism defined in Eq.~\eqref{err_prop}.

The projection is not entirely trivial, as the observables $H$ and $D_\mathrm{A}$ do not exhibit explicit dependence on some parameters in $\boldsymbol{\theta}$ (in particular they only depend on $\Omega_\mathrm{m,0}$ and $h$), leading to potentially ill-defined or numerically vanishing derivatives. This issue can be overcome by reducing the Fisher matrix to a smaller subspace involving only $\Omega_{\mathrm{m},0}$ and $h$, prior to the Jacobian transformation.

Due to numerical instabilities during the computation of the derivatives via finite differences, the resulting Fisher matrix ended up being non-positive definite. However, this is not the only issue. In fact, a more critical problem is the high degree of correlation between some parameters, which can lead to some eigenvalues becoming extremely small, making the matrix inversion numerically unstable. 
Nevertheless, we have already explained how to address this by applying the regularization technique from Eq.~\eqref{fish_curing}. 

In the regularization process, we need to be extremely cautious when choosing the regularization parameter $\alpha$. An overly aggressive choice for $\alpha$ can artificially change the uncertainties and distort the orientation of the confidence ellipses, whereas a very small value may fail to make the Fisher matrix positive-definite. 

To ensure a conservative and physically meaningful choice, we rely on two diagnostics. First, the regularized Fisher matrix must conserve the overall structure of the original likelihood, which would lead to statistically consistent iso-likelihood contours. Secondly, we compute the chi-squared per degrees of freedom (dof), $\chi^2/\mathrm{dof}$, for the mock data, using the $\chi^2$ and the assorted covariance matrix in the $\boldsymbol{\phi}$ basis, defined below. A well-chosen $\alpha$ must lead to $\chi^2/\mathrm{dof}\sim1$, which indicates that the model describes the data within the expected level of statistical fluctuations. For instance, if values are significantly different from unity, this would imply either an underfitting ($\chi^2/\mathrm{dof}\gg1$) or overfitting ($\chi^2/\mathrm{dof}\ll1$) regime.\footnote{Following these two criteria, we find, via trial and error, that the optimal value is $\alpha \simeq 3\cdot10^{11}$. The reason why $\alpha$ is so large is that the ``problematic'' eigenvalues, i.e. the ones that are negative, are quite small and close to machine precision, of the order of $\sim -10^{-16}$, while the proper positive eigenvalues are of the order of a few thousands. Thus, the product of the two (the value of $\alpha$ and the eigenvalues), is sufficiently small to not affect the $\chi^2$, while canceling the negative eigenvalues at the same time.} Finally, using the criterion that $\chi^2/\mathrm{dof}\sim1$ also ensures that our analysis is robust (in the sense of avoiding either overfitting or underfitting) and not affected by the particular choice of $\alpha$. We have also tested that slightly perturbing the value of $\alpha$ does not affect the results.

Once we have the Fisher matrix in the $\boldsymbol{\phi}=\{H(z_i),D_\mathrm{A}(z_i)\}$ basis, we can proceed with the consistency tests mentioned in section \ref{sec::consist_sect}. Each realization is reconstructed using the code \texttt{Genetic Algorithms}\footnote{\url{https://github.com/snesseris/Genetic-Algorithms}}, which is trained to minimize the chi-squared defined as 
\begin{equation}\label{eq:chi2_def}
    \chi^2=(\mathbf{X}_{\text{mock}}-\mathbf{X}_{\text{GA}})^T\;\mathbf{F}(\phi)\;(\mathbf{X}_{\text{mock}}-\mathbf{X}_{\text{GA}})\;,
\end{equation}
where $\mathbf{X}_\text{mock}=[H(z_1),D_\mathrm{A}(z_1),\,\ldots\,,H(z_{11}),D_\mathrm{A}(z_{11})]$ corresponds to the mock realizations and $\mathbf{X}_{\text{GA}}$ denotes the GA reconstructed values. The reconstruction is fully non-parametric and relies on the statistical information encoded in the Fisher matrix, without assuming any functional form for $H(z)$ or $D_\mathrm{A}(z)$, or even imposing Eq.~\eqref{cov_distance} so as to be able to test for non-FLRW universes, providing smooth, analytic functions for both. Then, given these two functions, we can proceed to reconstruct the null tests by using Eqs.~\eqref{eq:om_statistic}, \eqref{eq:ok_test} and \eqref{global_shear2}.

\subsection{Genetic algorithms}
Genetic algorithms (GA) are a class of unsupervised ML technique inspired by the principles of natural selection and biological evolution \cite{Arjona:2019fwb}. By emulating the evolutionary process (through selection, crossover and mutation) GAs iteratively optimize a population of candidate solutions to fit the given data. They are particularly well-suited for high-dimensional and non-differentiable optimization problems \cite{GA_review}. 

The algorithm begins by initializing a random population of candidate solutions, called chromosomes, using a basis of functions called the grammar and a set of operations: the usual numerical operations: $[+, -, \times, \div ]$, along with the genetic operations of crossover and mutation (discussed later on). In the context of regression, each of these chromosomes represent a set of parameters that define a trial function. These candidates are then evaluated based on the fitness function.

Following the fitness evaluation, the algorithm performs a selection process to determine which individuals will contribute to the next generation, favoring those with the best fitness. The selection procedure can be done following two different criteria: the roulette wheel selection,  which assigns reproduction probabilities proportional to their fitness, or tournament selection (which we use in this work), where each individual competes and only the best performers are chosen. In practice,  every member, i.e. function, in the GA population has its $\chi^2$ compared with another during the tournament selection, albeit also a stochastic component as the tournament is applied on a subset (random sample) of the best and worst (in terms of $\chi^2$) function. To be more clear, only a subset of the GA population of functions enters the tournament.

The selected individuals then undergo a reproduction process. There are two main operators: crossover or recombination, where two parents' chromosomes combine to perform an offspring, and mutation, stochastic alternation of the parameters (genes) of the chromosomes. In practice, the crossover joins different parts of members of the population to create new ``individuals", see for example Fig. 1 in Ref.~\cite{Nesseris_2012}. For example, two functions $f_1(x)=1+x+x^2+x^3$ and $f_2(x)=2+\sin(x)$, might be combined to give $g_1(x)=1+x+\sin(x)$ and $g_2(x)=2+x^2+x^3$, where the parts $1+x$ and $\sin(x)$ where picked from $f_1(x)$ and $f_2(x)$ to give $g_1(x)$ respectively, while $x^2+x^3$ and $2$ where picked from $f_1(x)$ and $f_2(x)$ to give $g_1(x)$ respectively.

Moreover, offspring are created by crossover replacing members that failed in the tournament. Then, all members have a small chance, determined by the mutation rate (a hyperparameter of the GA, typically set to $\sim 0.2$ or $0.3$ to be mutated, i.e. parts of their expression stochastically changed, e.g. $x^2\rightarrow x^3$ etc \cite{Nesseris_2012}.

The cycle of evaluation, selection, and reproduction is performed iteratively over multiple generations until the termination criterion is satisfied, i.e. a maximum number of generations (set to 500 in this work to ensure convergence of the GA) or the convergence of the populations \cite{Kamerkar_2023}. The GA also has some hyperparameters typically taken to be the population size, the mutation rate, the crossover rate, and the random seed number. In our case however, we found good convergence after fixing the population size to 100, the crossover rate to 0.75, the mutation rate to 0.3, and varied the random seed for five GA chains, ensuring all of them had converged and had a better $\chi^2$ than the fiducial model. On the other hand, we kept the complexity of the GA functions fixed to a length of four to avoid overfitting, as discussed in Ref.~\cite{Orjuela-Quintana:2022nnq}. In Fig.~\ref{flowcharts} we show a flowchart of the iterative process followed by the GA, summarizing a typical run of the algorithm.

Analogous to parametric approaches, we can define a likelihood function, $\mathcal{L}\propto e^{-\chi^2/2}$, to quantify the quality of the fitness statistically, where the $\chi^2$ is given by Eq.~\eqref{eq:chi2_def}. Thus, for a given chromosome $c$, we define a likelihood $\mathcal{L}(c)$. To perform proper statistical inference, the likelihood must be normalized by integrating over all possible functions that can encountered by the GA, which can be done by means of a path integral \cite{Nesseris_2012}. Then, once the error on $H(z)$ and $D_\mathrm{A}(z)$ can be reconstructed, the error on the null tests can be estimated via standard error propagation, following Ref.~\cite{Arjona:2020kco}. Furthermore, this approach can be generalized to handle correlated data to account for the covariance structure of the dataset. The confidence intervals can be computed by integrating over the normalized likelihood function of the best-fit function around $\pm1\sigma$, i.e. the uncertainty associated with the reconstruction, under the assumption of gaussianity/normality. The errorbands produced by the GA were tested against bootstrap MCMCs in Ref.~\cite{Nesseris_2012} (see Section II.C) and found to be in excellent agreement.

One of the main advantages of GAs is that, they do not rely on specific theoretical models, making them well-suited for model-independent reconstructions. This feature makes them particularly useful for performing null tests \cite{Nesseris_2022,Nesseris_2010} or reconstructing physical quantities in a theory agnostic manner \cite{Arjona:2019fwb,Aizpuru_2021}. Finally, the GA is a stochastic algorithm, meaning that every run with a different seed number will produce slightly different function, thus individual reconstructions are not interesting on their own and later on we choose to show an example reconstruction, but they are just a way to compress the data.

\subsection{Neural network pipeline}
In addition to these consistency tests, we explore the \emph{implicit} model dependence of the Fisher matrix on the cosmological background parameters and test the performance of the NN to discriminate the models ($\Lambda$CDM vs $w$CDM).

\begin{figure}[!t]
\centering
\begin{minipage}{0.495\textwidth}
\centering
\includegraphics[width=\textwidth]{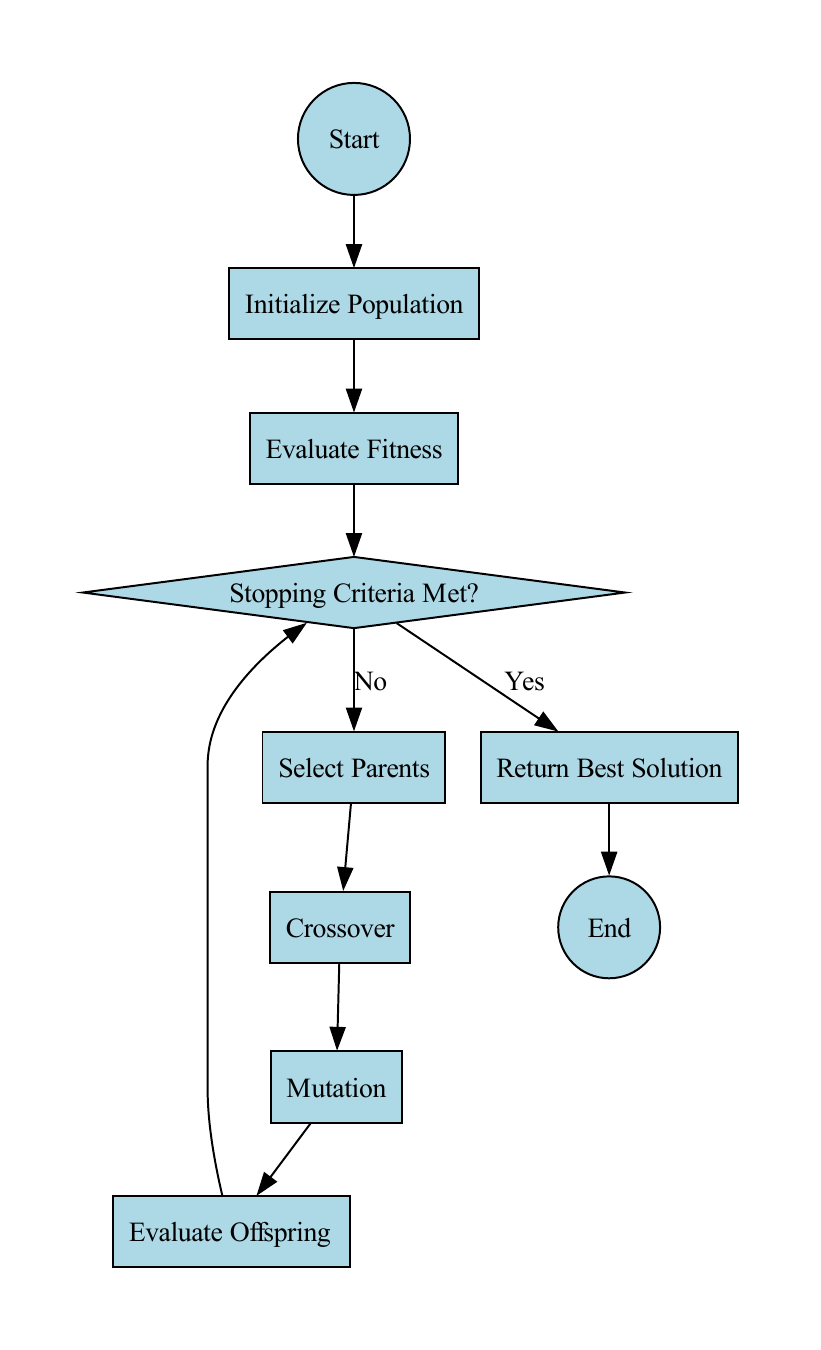}
\end{minipage}
\vspace{0em} 
\begin{minipage}{0.495\textwidth}
\centering
\includegraphics[width=\textwidth]{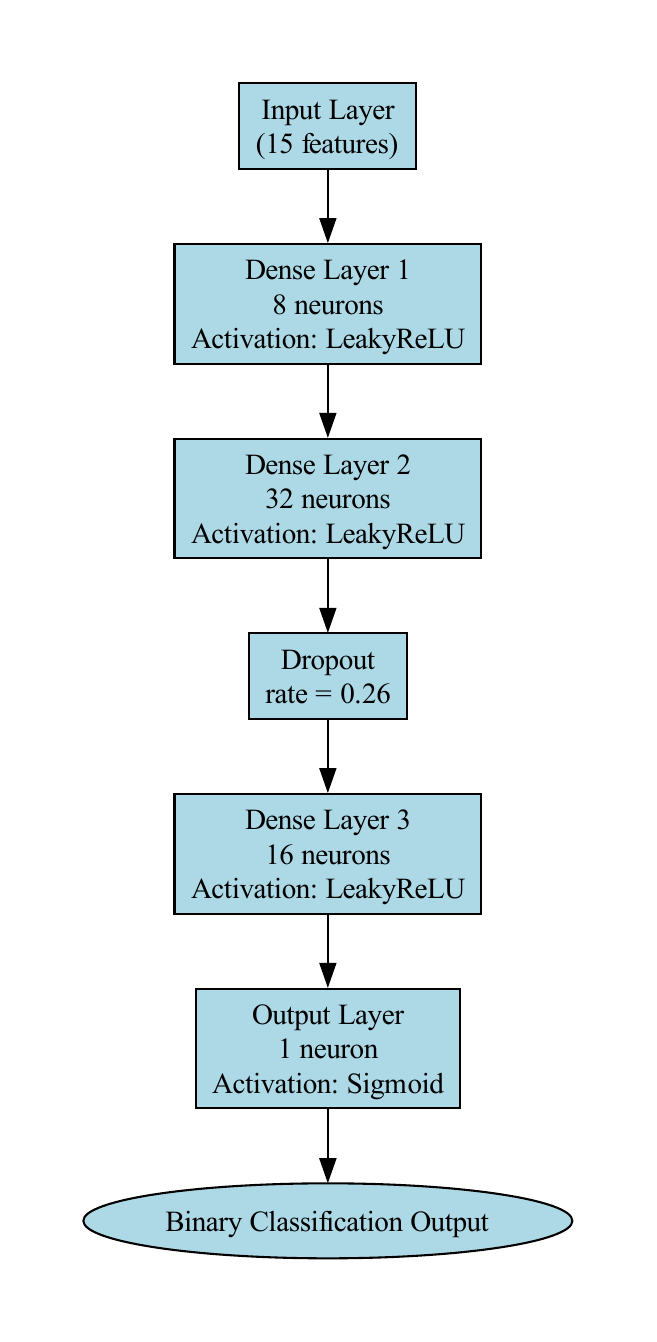}
\end{minipage}
\caption{Flowchart of the iterative processes followed by the GA (left) and the NN architecture (right), as implemented in our analysis.}
\label{flowcharts}
\end{figure}

For the classifier, we have made use of the Fisher matrix derived from the cosmological set of parameters $\boldsymbol{\theta}$ (in contrast to the consistency tests, where we used the $\boldsymbol{\phi}$ basis). Specifically, we fit the NN with the raw simulated fisher matrices in the $\theta$ basis (as we are interested in detecting the parameter dependence), but we use the $\phi$ basis for the null tests (as we are interested in the errors of the distances).

In this space dimension of the $\boldsymbol{\theta}$ basis, the Fisher matrix has dimension $5\times5$, but taking advantage of its symmetry, we can consider only the upper triangular components as features, resulting in a feature vector of 15 unique values. In order to pass on to the NN matrices of the same dimensions, we have to marginalize over $w$ so that all the matrices have $5\times5$ entries. While this marginalization over $w$ might lower the amount of information in the Fisher matrix, this was in fact our goal with this test.

We have generated 2,000 instances, 1,000 instances for each $\Lambda$CDM and $w$CDM cosmological models. Each sample has been labeled accordingly, with 1 assigned to $\Lambda$CDM and 0 to $w$CDM. These Fisher matrices have been computed over the range $0.2\leq\Omega_{\mathrm{m},0}\leq0.4$ and $-2\leq w\leq -1/3$, where all cases corresponding to $w=-1$ have been classified as $\Lambda$CDM. The values of $\Omega_\mathrm{m,0}$ and $w$ have been drawn randomly, assuming a uniform distribution in the aforementioned ranges, while the other parameters remain fixed following the fiducial values in Table \ref{fid_params}. In this way, the sampling covers a representative cosmological parameter space. 

The classification task has been performed with a NN built using \texttt{Tensorflow Keras} \cite{tf_keras}. To optimize and choose the most appropriate architecture, we have used \texttt{Optuna} \cite{Optuna_ref}, an optimization framework to find the optimal number of layers, neurons and hyperparameter values. This framework works by training the same dataset multiple times with different architectures and selecting those with the best accuracy performance.

Specifically, we have used a Tree-structured Parzen Estimator which implements a Bayesian optimization strategy to efficiently explore the hyperparameter space. The optimization was carried out over 150 independent trials, each corresponding to the training of an individual model with distinct hyperparameter configurations.

The optimized architecture found, consists of three hidden layers with 8, 32 and 16 neurons respectively, with a \texttt{LeakyRELU} activation function. The second hidden layer has a dropout rate of 0.265, whereas the first and third hidden layers are free of any regularizer. The input layer consists of 15 neurons, corresponding to the features, and the output layer has only one neuron with a sigmoid activation function. The weights and biases have been optimized during training using an Adam optimizer with a learning rate $\eta=0.0015$ and a binary cross-entropy loss function. Figure \ref{flowcharts} (right panel) shows a schematic representation of the architecture deployed in our analysis. 

Lastly, for model evaluation, we perform a $k$-fold stratified cross-validation \cite{kfoldCV}. This involves partitioning the dataset into $k$ equally sized subsets or folds. The model is trained and tested $k$ times, each time using a different fold as a test set and $k-1$ folds for training. The performance is then evaluated by means of the accuracy and confusion matrix metrics. 

\section{Results and discussion}
Following the approach described in Section \ref{meth_sect}, we now present the results of our analysis. We begin by showing the Fisher forecast uncertainties on the angular diameter distance $D_\mathrm{A}(z)$ and Hubble parameter rate $H(z)$. Next, we present the outcome of the consistency tests under the $\Lambda$CDM framework. Finally, we evaluate the performance metrics of the NN trained to distinguish cosmological models, evaluating its accuracy and robustness. 

\subsection{Fisher forecasts on $D_\mathrm{A}(z)$ and $H(z)$}
\begin{figure}[!t]
\centering
\begin{minipage}{0.495\textwidth}
\centering
\includegraphics[width=\textwidth]{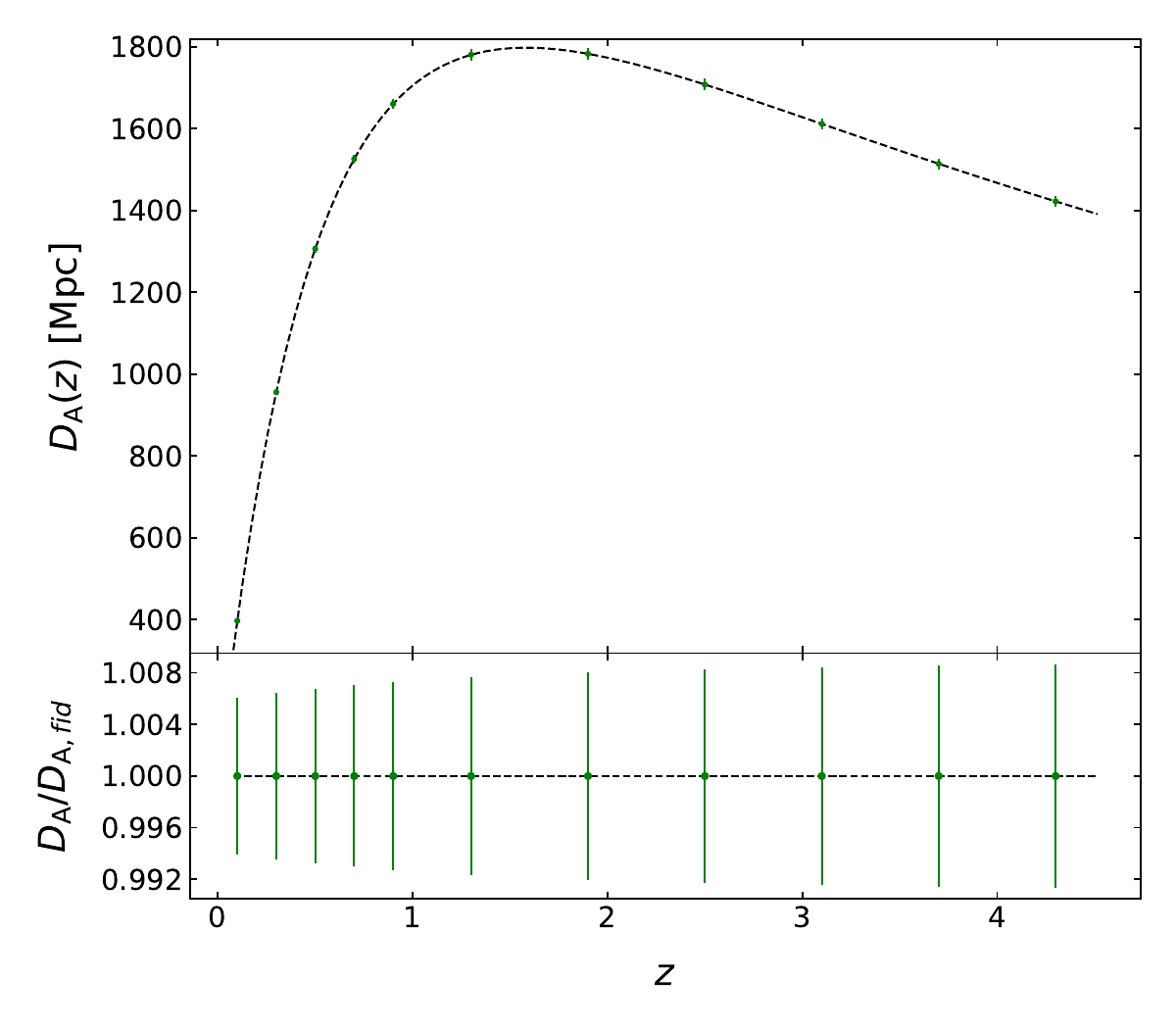}
\end{minipage}
\vspace{0em} 
\begin{minipage}{0.495\textwidth}
\centering
\includegraphics[width=\textwidth]{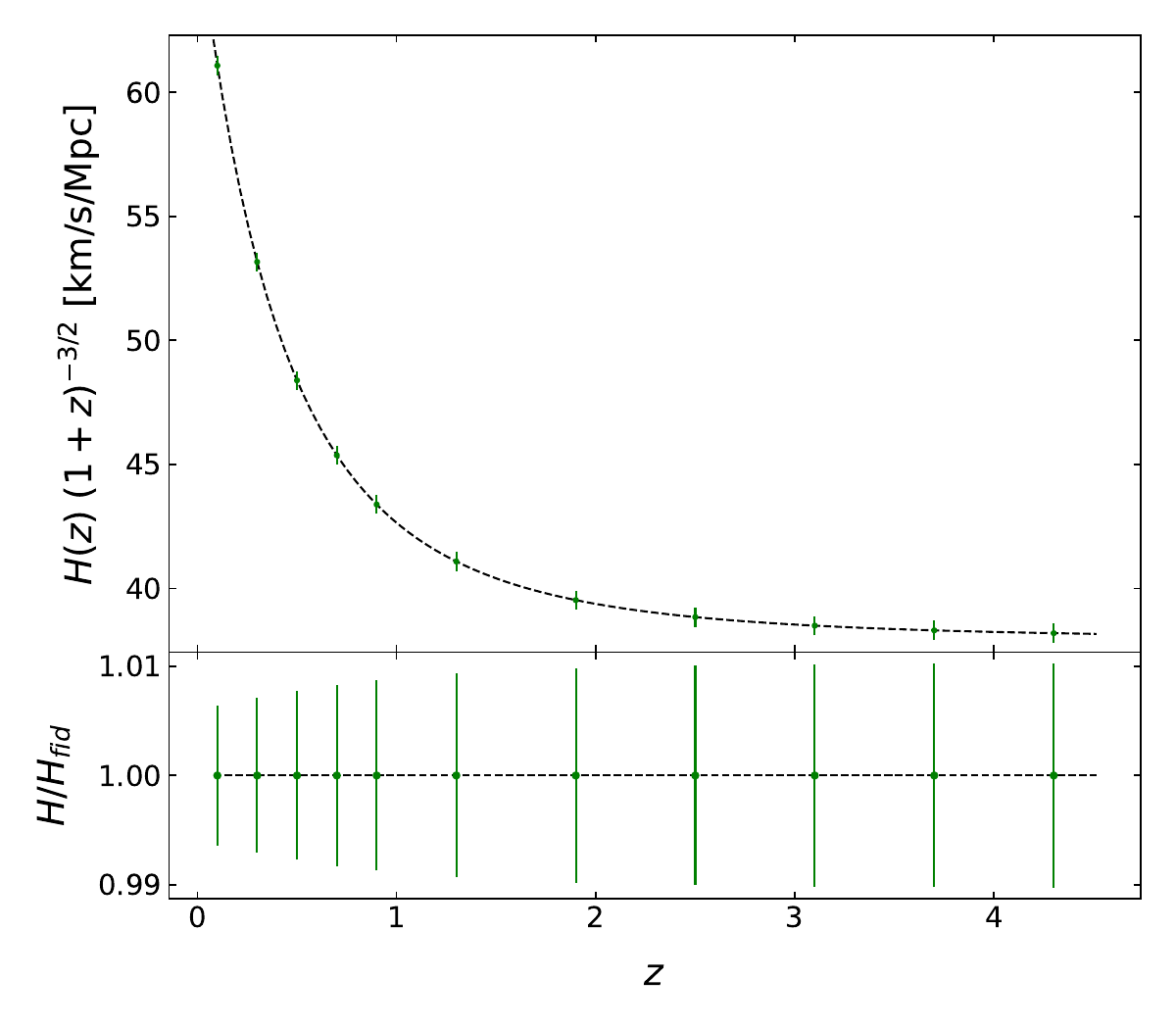}

\end{minipage}
\caption{We show the theoretical values of the angular diameter distance (left) and the Hubble parameter over $(1+z)^{3/2}$ (right) as functions of redshift (black dashed lines), alongside the theoretical values of Table~\ref{fid_params} (green points with error bars). In the bottom of each panels we show the values normalized to unity, with respect to the fiducial model.}
\label{HDA_fig}
\end{figure}

We now present the forecasted 1$\sigma$ uncertainties on the angular diameter distance $D_\mathrm{A}(z)$ and $H(z)$. In Fig.~\ref{HDA_fig} we show the predicted uncertainties at the discrete redshift bins (highlighted in green), overlaid on the corresponding theoretical curves for each observable (black dashed line). For the Hubble parameter (right panel), we plot the rescaled quantity $H(z)\,(1+z)^{-3/2}$ in order to factor out the dominant matter domination redshift dependence, effectively flattening the curve and allowing a clearer visual interpretation of the forecasted uncertainties. 

As can be seen, the expected uncertainties are small relative to the amplitude of the cosmological observables, indicative of the high statistical precision of SPHEREx. The errors increase modestly with the redshift, as expected due to the decreasing galaxy number density. Then, in Table~\ref{tab:constraints} we quantify these forecasted relative uncertainties on $D_\mathrm{A}(z)$ and $H(z)$ across the full set of eleven redshift bins. The relative error on $D_\mathrm{A}(z)$ remains below approximately $0.87\%$ over the entire redshift range, while the uncertainties on $H(z)$ increase more noticeably, reaching up to $\sim1\%$ at $z=4.3$. 

\begin{table}[!t]
\centering
\renewcommand{\arraystretch}{1.2}
\resizebox{\textwidth}{!}{%
\begin{tabular}{|c|ccccccccccc|}
\hline
$z$ & 0.1 & 0.3 & 0.5 & 0.7 & 0.9 & 1.3 & 1.9 & 2.5 & 3.1 & 3.7 & 4.3 \\ \hline
$\sigma[D_\mathrm{A}]/D_\mathrm{A}$ [\%] & 
0.6089 & 0.6423 & 0.6739 & 0.7019 & 0.7262 & 0.7642 & 0.8023 & 0.8268 & 0.8436 & 0.8557 & 0.8648 \\ \hline
$\sigma[H]/H$ [\%] & 
0.6385 & 0.7072 & 0.7718 & 0.8265 & 0.8704 & 0.9309 & 0.9792 & 1.0026 & 1.0149 & 1.0218 & 1.0260 \\ \hline
\end{tabular}}
\caption{Forecasted 1$\sigma$ relative uncertainties (\%) on the angular diameter distance $D_\mathrm{A}(z)$ and Hubble parameter $H(z)$ for SPHEREx, over 11 redshift bins.}
\label{tab:constraints}
\end{table}

To assess the internal consistency of these results, we perform a $\chi^2$ minimization, which quantitatively compares the forecasted uncertainties against the corresponding theoretical model predictions. The minimization yields optimal estimates of $\Omega_{\mathrm{m},0}=0.319$ and $h=0.668$, in close agreement with the fiducial cosmology initially assumed, thus validating our analysis. 

\subsection{$\Lambda$CDM consistency test results}
\begin{figure}[!t]
    \centering
    \includegraphics[width=0.325\textwidth]{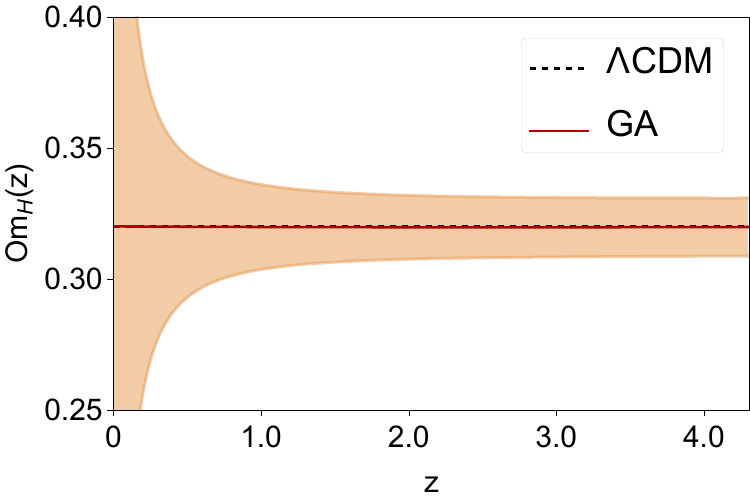}
    \includegraphics[width=0.325\textwidth]{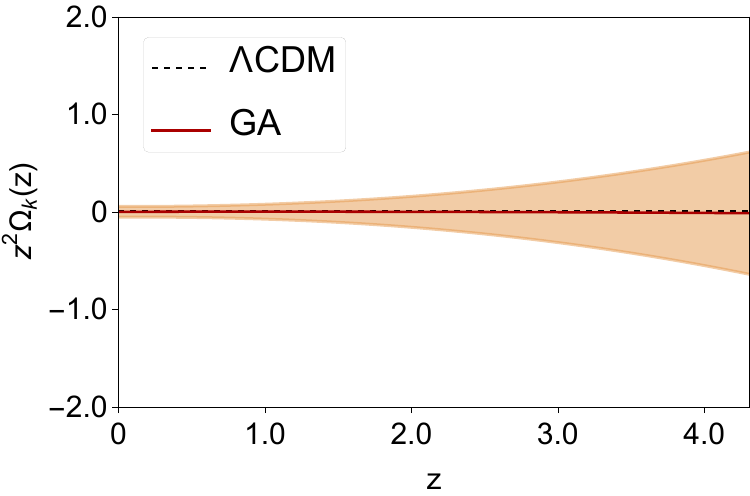}
    \includegraphics[width=0.325\textwidth]{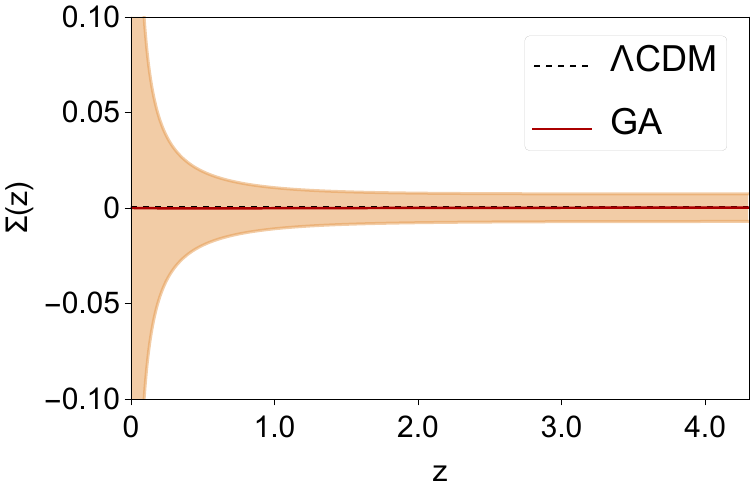}
    \caption{GA reconstructions of the consistency tests mentioned in Section \ref{sec::consist_sect}. The black dashed line corresponds to the best-fit $\Lambda$CDM model, the red line is the GA fit and the orange shaded region corresponds to the 1$\sigma$ GA uncertainties.}
    \label{null_fig}
\end{figure}

In Fig.~\ref{null_fig} we show the two reconstructed null tests: the $O\mathrm{m}_\mathrm{H}$ (left panel), $z^2\,\Omega_k$ (middle panel) diagnostic and the $\Sigma(z)$ (right panel). The best-fit $\Lambda$CDM model corresponds to the dashed black line, whereas the red solid line and shaded orange region correspond to the GA best-fit and $1\sigma$ error respectively. 

In all three diagnostics, we find that the GA reconstructions remain consistent with the null hypothesis within the $1\sigma$ uncertainty level. That is, there is no statistical deviation from the homogeneous and isotropy assumptions as expected, given our fiducial model is the $\Lambda$CDM model.

Thus, our analysis demonstrates that we are able to constrain the uncertainties of the $O\mathrm{m}_\mathrm{H}$ test within a relative error of approximately $\pm6\%$ for redshift values greater than $z\sim 1$. Nevertheless, the uncertainties tend to increase as $z\to0$, which is expected due to the $\sim1/z$ dependence, as discussed in Section \ref{sec::consist_sect}. 

On the other hand, the $z^2\Omega_k$ diagnostic exhibits low uncertainties at low redshift values, as the $z^2$ factor regularizes the divergence of $\sim1/z^2$. However, this same $z^2$ factor causes the absolute uncertainties to grow with redshift, since any relative errors in $z^2\Omega_k$ are amplified. Despite this increase, the test remains statistically consistent with zero within the forecasted $1\sigma$ errors, i.e. no significant deviation from spatial flatness and homogeneity of the Universe. 

Similarly to the $O\mathrm{m}_\mathrm{H}$ test, the $\Sigma(z)$ diagnostic exhibits relatively small uncertainties at intermediate and high redshifts, while diverging at low redshift. Despite this low-redshift divergence, the $1\sigma$ uncertainties remain consistent with zero, indicating no significant deviation from large-scale spatial isotropy in the Universe.  

Our results are in good agreement with those reported by Ref.~\cite{Nesseris_2022}, who similarly employed GA reconstructions methods to forecast $O\mathrm{m}_\mathrm{H}$, $z^2\Omega_k$ and $\Sigma$ tests with \textit{Euclid} mock data. In line with their results, we obtain uncertainties at few-percent-level in both tests across wide redshift range. Their analysis exhibits similar qualitative behavior, showing consistency with the $\Lambda$CDM expansion history and indicating no statistical deviation from spatial flatness, but also the more agnostic nature of the GA. 

It is important to emphasize that, although the mock data used in our analysis possesses relatively small uncertainties, at the level of $\sim1\%$ at maximum, the reconstructed quantities derived, i.e. $O\mathrm{m}_\mathrm{H}$, $z^2\Omega_k$ and $\Sigma$, exhibit significantly larger uncertainties. This discrepancy is primarily driven by error propagation through non-linear transformations of $H(z)$, $r_\mathrm{c}(z)$, their derivatives, and the presence of redshift dependent structures that amplify numerical sensitivity. 

\subsection{Neural network classification results}
To assess the performance of the NN classifier, we have implemented a $k$-fold stratified cross-validation, explained in Section \ref{meth_sect}, with $k=5$. For each fold, the data was split into an 80\% - 20\% training-test partition, and the training subset was further split into an 80\% - 20\% training-validation subsets.

After training, the predictions were generated for the test set. For each fold, we computed the accuracy and normalized confusion matrix, quantifying the average performance and variance of the classifier across independent data partitions. We show in Appendix~\ref{app::loss_acc} the loss and accuracy curves for the fifth fold. We observe that the model achieves a validation accuracy exceeding 95\% and a validation loss below 1\%. The performance trends for the remaining folds exhibit consistent behavior. 

\begin{figure}[!t]
\centering
\includegraphics[scale=0.65]{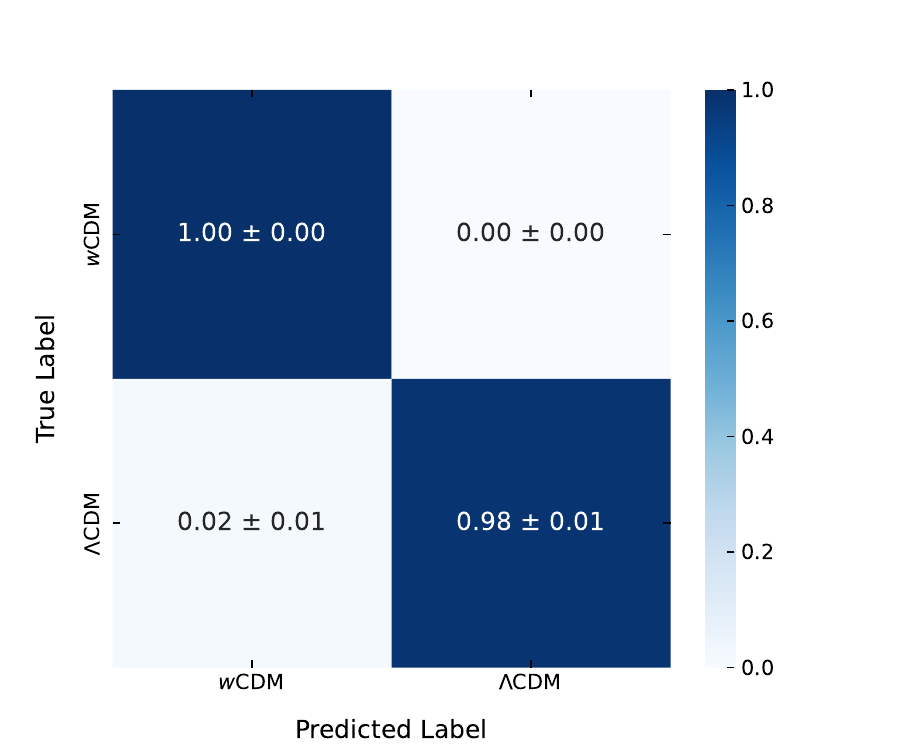}
\caption{Confusion matrix results from the $\Lambda$CDM - $w$CDM classification. The top left and bottom right correspond to the true negatives (TN) and true positives (TP), representing correct classifications. The off-diagonal terms represent the false positives (FP) and false negatives (FN).}
\label{conf_mat_pic}
\end{figure}

The mean accuracy across all folds is $0.98\pm0.01$, demonstrating its ability to generalize and distinguish between these two cosmological models considered. 
Further insight is gained by examining the confusion matrix, shown in Fig.~\ref{conf_mat_pic}, which summarizes the classifier predictions in detail. Notably, all samples belonging to the $w$CDM model were correctly classified without any mislabeling. In contrast, a small fraction of $2\%$ of the $\Lambda$CDM instances were misclassified as $w$CDM, suggesting a small overlap in the feature space. Nevertheless, this low misclassification rate underscores the effectiveness of the classifier. 

These results show that there is a lot of cosmological information encoded in Fisher matrices, retaining sufficient sensitivity to the underlying physical parameters of the models under consideration. This enables the classifier to discriminate between two cosmological scenarios, i.e. $\Lambda$CDM and $w$CDM, based solely on the statistical information encoded in Fisher matrices. 

Although these results highlight the potential of this method for future classification tasks, they also reveal a physically significant output: while it is commonly assumed that the cosmological dependence of the covariance matrix is weak, our analysis shows that a NN can reliably distinguish between $\Lambda$CDM and $w$CDM models based solely on this input. This suggests that the covariance and Fisher matrix retain subtle but non-trivial imprints of the underlying cosmological model. Therefore, this test not only validates the classification pipeline, but challenges the assumption of weak model dependence in $F_{ij}$. 

\section{Conclusions}
In this work, we have leveraged Fisher matrix forecasting for the upcoming SPHEREx survey to constrain cosmological observables, test some of the fundamental assumptions of homogeneity and isotropy of the $\Lambda$CDM model, and challenge the assumption of weak model dependence in the Fisher matrix using machine learning. 

Our Fisher forecast analysis shows that the angular diameter distance $D_\mathrm{A}(z)$ and the Hubble parameter $H(z)$ can be predicted from the BAO data with high accuracy. The relative uncertainties for $D_\mathrm{A}(z)$ remain below approximately 0.87\%, while we reach nearly $1\%$ for $H(z)$. The internal consistency of these constraints is supported by $\chi^2$ minimization, which recovers the initial cosmology for $\Omega_{\mathrm{m},0}$ and $h$, albeit a small caveat of our analysis is the regularization process, as one needs to be extremely cautious when choosing the regularization parameter $\alpha$. However, in this particular case we have performed extensive tests to make certain our analysis is robust.

Using the forecasted data, we reconstructed the $O\mathrm{m}_\mathrm{H}$, $z^2\,\Omega_k$ and $\Sigma$ null tests of $\Lambda$CDM. All three diagnostics are consistent with the null hypothesis at the $1\sigma$ uncertainty level. Notably, the $O\mathrm{m}_\mathrm{H}$ test achieves uncertainties of almost 6\% for redshift values larger than $z\sim1$. However, its error diverges at low redshift due to the $\sim1/z$ dependence. In contrast, the curvature test maintains low uncertainties at low redshift, although, it exhibits greater uncertainties at larger redshifts because of the $z^2$ factor. Despite this, it remains consistent with the zero curvature. 

Finally, we applied a NN binary classifier for $\Lambda$CDM and $w$CDM models using their Fisher matrices, after marginalization over the EoS parameter $w$ in the $w$CDM case, as features. It has demonstrated an accuracy of $0.98\pm0.01$, correctly identifying all of the $w$CDM samples and misclassifying only 2\% of $\Lambda$CDM cases. These results indicate that, contrary to the common assumption that Fisher matrices are weakly model-dependent, subtle imprints of the cosmological model remain encoded in its structure. This suggests that Fisher matrices contain more model-specific information than previously assumed, and that machine learning methods can effectively extract and leverage these differences for cosmological model discrimination. 

In summary, our work demonstrates the potential of the upcoming SPHEREx survey to constrain cosmological observables and test the internal consistency tests of the $\Lambda$CDM model. In particular, we find that SPHEREx will be capable of detecting potential deviations from homogeneity and isotropy at the percent level, thus allowing for very precise and accurate consistency tests of $\Lambda$CDM model, opening the door for searches of physics beyond the standard cosmological model. 

Lastly, we have also tested the long-held assumption that the Fisher matrix only depends very weakly on the underlying fiducial model, thus in principle not affecting cosmological inference. However, we have shown that using a NN it is in fact possible to discriminate different models, based solely on the covariance matrix of the data, thus challenging the aforementioned assumption. Thus, this result highlights the need for more accurate analyses given the exquisite data of current surveys, in order to maximize their potential.

\section{Acknowledgments}
We would like to thank O.~Dor\'e and D.~Sapone for useful discussions. SN and IO acknowledge support from the research project PID2021-123012NB-C43 and the Spanish Research Agency (Agencia Estatal de Investigaci\'on) through the Grant IFT Centro de Excelencia Severo Ochoa No CEX2020-001007-S, funded by MCIN/AEI/10.13039/501100011033. IO is also supported by the fellowship LCF/BQ/DI22/11940033 from ``la Caixa” Foundation (ID 100010434).

\section{Code and data availability}
All code and scripts developed for this analysis will be available, upon publication, at \url{https://github.com/amata21/LCDM_null_tests_SPHEREx}, while the relevant data products will be available at \url{https://zenodo.org/records/16160211}. 

\appendix
\section{Loss and accuracy plots} \label{app::loss_acc}
\begin{figure}[!t]
    \centering
    \includegraphics[scale=0.47]{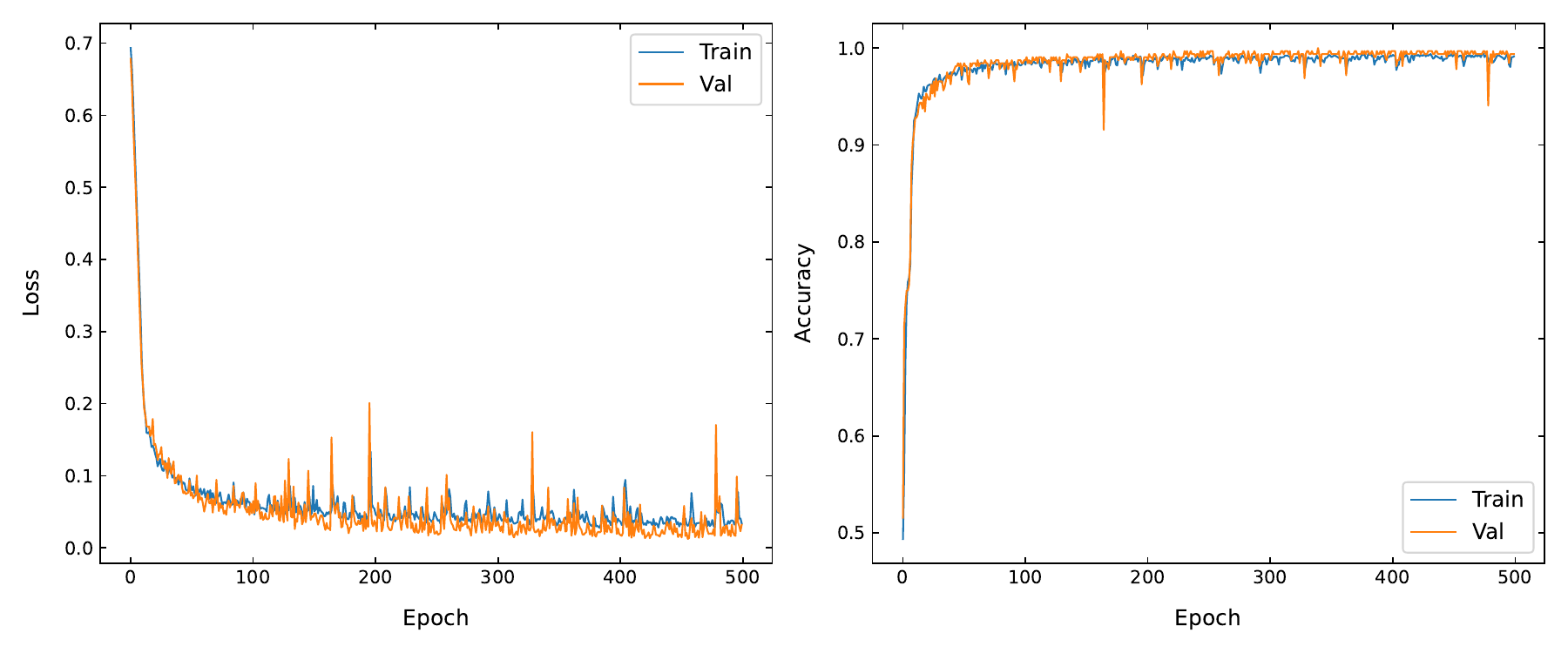}
    \caption{Loss (left) and accuracy (right) curves for both the training (blue) and validation (orange) data for the NN model implemented.}
    \label{AL_plot}
\end{figure}

For completeness, in Fig.~\ref{AL_plot} we present the loss and accuracy plots for the training and validation sets of  the NN model. 

\newpage

\bibliography{biblio}

\end{document}